\documentclass[conference]{IEEEtran}
\usepackage[cmex10]{amsmath}
\usepackage{graphicx}
\usepackage{epstopdf}
\usepackage{cite}
\usepackage{algorithm} 
\usepackage{algorithmic} 
\usepackage{stfloats}

\begin{document}
\title{Optimal Pilot Symbols Ratio in terms of Spectrum and Energy Efficiency in Uplink CoMP Networks}
\author{\IEEEauthorblockN{Yuhao~Zhang, Qimei~Cui, and Ning~Wang}
\IEEEauthorblockA{School of Information and Communication Engineering,\\
Beijing University of Posts and Telecommunications, Beijing, 100876, China\\
Email: cuiqimei@bupt.edu.cn}}

\maketitle

\begin{abstract}
In wireless networks, Spectrum Efficiency~(SE) and Energy Efficiency~(EE) can be affected by the channel estimation that needs to be well designed in practice. In this paper, considering channel estimation error and \mbox{non-ideal} backhaul links, we optimize the pilot symbols ratio in terms of SE and EE in uplink Coordinated Multi-point~(CoMP) networks. Modeling the channel estimation error, we formulate the SE and EE maximization problems by analyzing the system capacity with imperfect channel estimation. The maximal system capacity in SE optimization and the minimal transmit power in EE optimization, which both have the closed-form expressions, are derived by some reasonable approximations to reduce the complexity of solving complicated equations. Simulations are carried out to validate the superiority of our scheme, verify the accuracy of our approximation, and show the effect of pilot symbols ratio.
\end{abstract}

\vspace{2mm}

\section{Introduction}\label{sec:1}
In Coordinated Multi-point~(CoMP) networks, Spectrum Efficiency~(SE) and Energy Efficiency~(EE), the important indexes in wireless networks~\cite{Ref4,Ref9}, can be both improved by the provided cooperative diversity gains~\cite{Ref12,Ref10}. For uplink transmission, Joint Reception CoMP~(JR-CoMP), the efficient uplink CoMP scheme, is characterized by simultaneous reception at multiple cooperative nodes with fully data and control information exchange~\cite{Ref2}. It is known that the closed-form approximation for EE-SE trade-off is derived with idealistic and realistic accuracy demonstration on the assumption that the backhaul links are prefect~\cite{Ref12}. However, ideal backhaul links and perfect Channel State Information~(CSI) are limited and impossible in practical networks~\cite{Ref8}. Therefore, we mainly study the SE and EE maximization based on non-ideal backhaul links and imperfect CSI.

In practice, only imperfect and non-real-time (long-term) CSI can be obtained and exchanged in the networks due to the channel estimation error and the capacity limited backhaul links. These limitations need to be well considered and have already catched much research attentions. In~\cite{Ref13}, considering the backhaul link constraint, different theoretical uplink CoMP concepts are analyzed. And the framework incorporating these concepts is provided with practical CoMP algorithms. In~\cite{Ref14}, the optimal clustering and rate allocation scheme for JR-CoMP networks with delayed CSI feedback is proposed by taking a stochastic decision approach. However, these works only assume non-ideal backhaul links and imperfect CSI sharing but without the consideration of channel estimation error.

It is known that the channel estimation is usually carried out by periodic pilot or training symbols transmission, where the estimation error occurs inevitably~\cite{ChannelError,Ref7,Ref3}. It is straightforward that too much and too less pilot symbols in channel estimation will both affect the system performance, so its suitable ratio needs to be determined rationally. In terms of SE and EE, decreasing transmit power and reducing pilot symbols will overcome system throughput deterioration, which unfortunately leads to the degradation of Bit Error Probability~(BEP) performance~\cite{ChannelError}. In order to find the optimal trade-off, the power and spacing of the pilot symbols are optimized to maximize SE in adaptive modulation OFDM networks with the imperfect CSI~\cite{Ref7}. And in~\cite{Ref3}, closed-form BEP expression is derived for Filter Bank Multicarrier~(FBM) networks, based on which the optimal power allocation between pilot and data symbols is proposed to minimize BEP. In uplink CoMP networks with perfect backhaul links, the channel estimation error is modeled and the system-level computer simulation is conducted to investigate system and user throughput~\cite{Ref15}. However, to the best of our knowledge, the pilot symbols assignment for optimal SE and EE in JR-CoMP networks with non-ideal backhaul links has never been discussed so far.

In this paper, we investigate optimal pilot symbols ratio for channel estimation to maximize SE and EE in uplink CoMP networks with non-ideal and limited backhaul links. Modeling the channel estimation error, we first formulate the SE and EE maximization problems by deriving the system capacity. Utilizing the optimal derivative conditions in SE maximization and the Lagrange multiplier method in SE maximization, the optimal pilot symbols ratio at every cooperative node are derived, based on which the maximal system capacity in SE optimization and the minimal transmit power in EE optimization are obtained consequently. To avoid solving complicated equations, we introduce some reasonable approximate relationships to get the closed-form expressions of these results. Finally, numerical simulations validate the superiority of our scheme and the accuracy of our approximation, and reveal that with the increase of pilot symbols ratio, SE and EE first rise rapidly in channel estimation limited region, and then decrease slowly in useful information limited region.

The rest of the paper is organized as follows. System model is described in Section~\ref{sec:2}. SE and EE optimizations are resolved in Section~\ref{sec:3} and Section~\ref{sec:4}. Simulations are provided in Section~\ref{sec:5}, followed by conclusion in Section~\ref{sec:6}.
\section{System Model}\label{sec:2}
In uplink \mbox{JR-CoMP} networks, a common \mbox{single-antenna} User Equipment~(UE) transmits information to $M$ coordinated Base Stations~(BSs), which can process the received signal cooperatively. These $M$ BSs are connected through non-ideal and limited backhaul links (\emph{e.g.}, wireless links), which cannot afford too much traffic loads. Therefore, with the consideration of backhaul exchange delay and in order to reduce backhaul overhead, only \mbox{long-term} CSI can be exchanged during long time interval among the $M$ coordinated BSs. Due to the limitation, only the \mbox{non-coherent} \mbox{JR-CoMP} can work in this situation. Moreover, the CSI is detected and estimated at each BS with errors, \emph{i.e.}, only imperfect CSI can be obtained. The flat block fading is assumed to characterize the complex channel gain between BS~$m$ and the common UE, denoted by $h_m$. $W$ is the system bandwidth and $n$ is the Additive White Gaussian Noise~(AWGN) with power $N=W \times N_0$, where $N_0$ is the Power Spectral Density~(PSD) of the noise.
\subsection{Signal Model}\label{sec:2:1}
In each time frame, the desired information $x$ is transmitted to the cluster of $M$ cooperative BSs by the UE with transmit power $P$. The received signal at BS~$m$ can be expressed as
\begin{equation}\label{ReceSignal_BS}
    y_m={\sqrt{P}{{h}_{m}}x}+i_m+n,
\end{equation}
where $E\{{{\left| {{x}} \right|}^2}\} = 1$. $i_m$, with the power $I_m$, is the overall interference at BS $m$ caused by other UEs, using the same resource, outside the $M$ coordinated BSs.

Therefore, with perfect synchronization among the cluster and considering the imperfect channel estimation, the \mbox{non-coherent} combined received signal can be expressed as
\begin{equation}\label{ReceSignal_real}
    y = \sum\limits_{m = 1}^M {\sqrt P {{\widehat h}_m}x}  + \sum\limits_{m = 1}^M {\sqrt P {{\widetilde h}_m}x}  + \sum\limits_{m = 1}^M {\left({i_m}+{n}\right)},
\end{equation}
where ${{\widehat h}_m}$ is the result of imperfect channel estimation, ${{\widetilde h}_m}$ is the corresponding error, and $h_m = {{\widehat h}_m} + {{\widetilde h}_m}$.
\subsection{Channel Estimation Error}\label{sec:2:2}
It is known that the channel estimation is based on pilot symbols detection followed by data demodulation in wireless networks~\cite{ChannelError}. The periodicity of pilot symbols depends on the channel coherence over time, frequency and space etc. It is assumed that pilot symbols are transmitted during the channel coherence interval, \emph{i.e.}, $L$ symbols frame in this paper.

We denote the ratio of the pilot symbols for channel estimation by $\alpha_m \ (0 \leq \alpha_m \leq 1)$ at BS $m$, therefore, $(1-\alpha_m)$ represents the other part for useful information. Under block fading channel, the Minimum Mean Square Error~(MMSE) of the estimation at BS $m$ can be expressed~\cite{ChannelError}, as given by
\begin{equation}\label{MMSE}
    {e_m} = \frac{1}{{1 + {\alpha _m} \cdot L \cdot {SNR}_m}},
\end{equation}
where ${SNR}_m$ is SNR of received signal at BS~$m$, given by
\begin{equation}\label{SNR_BS}
    {SNR}_m = \frac{{P{{\left| {{h_m}} \right|}^2}}}{I_m+N}.
\end{equation}

Hence, ${{\widetilde h}_m}$ and ${{\widehat h}_m}$ then can be formulated in terms of ${e_m}$, as written by
\begin{equation}\label{estimation_h}
    {|{{{\widetilde h}_m}}|^2} = {\left| {{h_m}} \right|^2} \cdot {e_m},\ \ {|{{{\widehat h}_m}}|^2} = {\left| {{h_m}} \right|^2} \cdot \left( {1 - {e_m}} \right).
\end{equation}
\subsection{System Capacity}\label{sec:2:3}
With imperfect channel estimation, ${S_m} = P{|{{{\widehat h}_m}}|^2}$ turns out to be the useful signal strength at BS $m$, while ${Z_m} = P{|{{{\widetilde h}_m}}|^2}$ degrades into the interference, which is treated as a noise for average performance due to the randomness of channel estimation each time in practice. Among the received power $S_m$, $\alpha_m \cdot S_m$ is devoted to pilot symbols demodulation, thus only $(1-\alpha_m) \cdot S_m$ is available for desired information. Therefore, using \eqref{ReceSignal_real} and \eqref{estimation_h}, the SNR for desired information can be obtained under \mbox{non-coherent} combination, as given by
\begin{equation}\label{SNR_origin}
    {SNR} = \frac{{\sum\limits_{m = 1}^M {\left( {1 - {\alpha _m}} \right)P{{\left| {{h_m}} \right|}^2}\left( {1 - {e_m}} \right)} }}{{\sum\limits_{m = 1}^M {P{{\left| {{h_m}} \right|}^2}{e_m}} + \sum\limits_{m = 1}^M {{I_m}} + M \cdot {N}}}.
\end{equation}

For simplicity and without loss of generality, it is assumed that the overall interference power $I_m$ is identical at each BS for average performance. Then substituting (\ref{MMSE}) and (\ref{SNR_BS}) into (\ref{SNR_origin}), the SNR is then transformed, as expressed by
\begin{equation}\label{SNR}
    {SNR} = \frac{{\sum\limits_{m = 1}^M {\left( {1 - {\alpha _m}} \right){SNR}_m\frac{{{\alpha _m} \cdot L \cdot {SNR}_m}}{{1{\rm{ + }}{\alpha _m} \cdot L \cdot {SNR}_m}}} }}{{\sum\limits_{m = 1}^M {\frac{{{SNR}_m}}{{1 + {\alpha _m} \cdot L \cdot {SNR}_m}}} + M}}.
\end{equation}
According to Shannon capacity formula, the achievable uplink data rate $C$ can be attained, as given by
\begin{equation}\label{coh_AchiRate}
    C = W \cdot {\log _2}\left( {1 + {SNR}} \right).
\end{equation}
\section{SE Optimization}\label{sec:3}
The SE is defined as the radio between the achievable uplink data rate and the system bandwidth, as given by
\begin{equation}\label{SE_indicator}
{\eta _S} = \frac{C}{W} = {\log _2}\left( {1 + {SNR}} \right).
\end{equation}

Due to monotonicity of $Log$ function, maximizing ${\eta _S}$ is equivalent to maximizing $SNR$. Thus, given fixed transmit power $P$, the SE optimization problem can be formulated as
\begin{equation}\label{SE_Opt}
    \mathop {\max }\limits_{\{\alpha _m\}} \ SNR \tag{\textbf{P1}},
\end{equation}
\begin{equation}\nonumber
    \rm{s.t.}\ \ 0 \le {\alpha _m} \le 1,\ m \in \{1, 2, \cdots, {M}\}.
\end{equation}

It can be proved that (\ref{SE_Opt}) is concave since its Hessian matrix is negative definite and the feasible region is linear, which guarantee the existence of exclusive maximum. Moreover, using this concave property, the optimal $\alpha _m$, denoted by $\alpha _m^*$, must satisfies $0<{\alpha _m^*}<1$ because $SNR > 0$ when $0<{\alpha _m}<1$ and $SNR = 0$ when ${\alpha _m} = 0$ or ${\alpha _m} = 1$.

\vspace{2mm}

\newtheorem{proposition}{\textbf{Proposition}}
\begin{proposition} \label{SE_proposition}
\textit{With fixed transmit power $P$ at the common UE, the approximate maximal system capacity $C^*$ in \mbox{non-coherent} \mbox{JR-CoMP} networks can be presented by}
\begin{equation}\label{Maximum_Capacity}
    C^* = 2 \cdot W \cdot {\log _2}\left( {{ {\frac{{ - {b_{SE}} + \sqrt {b_{SE}^2 + 4 \cdot M \cdot {c_{SE}}} }}{{2 \cdot M}}}}} \right),
\end{equation}
\textit{where}
\begin{equation}\nonumber
    {b_{SE}} = \sum\limits_{m = 1}^M {\frac{{2 \cdot {SNR}_m}}{{\sqrt {L \cdot {SNR}_m} }}},
\end{equation}
\begin{equation}\nonumber
    {c_{SE}} = \sum\limits_{m = 1}^M {\left( {\frac{{L \cdot {SNR}_m + 2}}{L} + 1} \right)}.
\end{equation}
\end{proposition}


\begin{IEEEproof}
The first derivative of $SNR$ can be expressed as
\begin{align}\label{df_SNR}
    \hspace{-1mm} \frac{{\partial {SNR}}}{{\partial {\alpha _m}}} & \hspace{-1mm} = \hspace{-1mm} - \frac{{L \hspace{-1mm} \cdot \hspace{-1mm} {SNR}_m^2 \hspace{-1mm} \left( {\alpha _m^2 \hspace{-1mm} \cdot \hspace{-1mm} L \hspace{-1mm} \cdot \hspace{-1mm} {SNR}_m \hspace{-1mm} + \hspace{-1mm} 2 \hspace{-1mm} \cdot \hspace{-1mm} {\alpha _m} \hspace{-1mm} - \hspace{-1mm} 1} \right)}}{{\left( {\sum\limits_{m = 1}^M \hspace{-1mm} {\frac{{{SNR}_m}}{{1 + {\alpha _m} \cdot L \cdot {SNR}_m}}} \hspace{-1mm} + \hspace{-1mm} M} \right) \hspace{-1mm} {{\left({1 \hspace{-1mm} + \hspace{-1mm} {\alpha _m} \hspace{-1mm} \cdot \hspace{-1mm} L \hspace{-1mm} \cdot \hspace{-1mm} {SNR}_m} \right)}^2}}} \nonumber \\
    & \hspace{-1mm} + \hspace{-1mm} \frac{{L \hspace{-1mm} \cdot \hspace{-1mm} {SNR}_m^2 \hspace{-1mm} \left( {\sum\limits_{m = 1}^M \hspace{-1mm} {\left( {1 \hspace{-1mm} - \hspace{-1mm} {\alpha _m}} \right){SNR}_m\frac{{{\alpha _m} \cdot L \cdot {SNR}_m}}{{1{\rm{ + }}{\alpha _m} \cdot L \cdot {SNR}_m}}} } \right)}}{{{{\left( {\sum\limits_{m = 1}^M \hspace{-1mm} {\frac{{{SNR}_m}}{{1 + {\alpha _m} \cdot L \cdot {SNR}_m}}} \hspace{-1mm} + \hspace{-1mm} M} \right)}^2}{{\left( {1 \hspace{-1mm} + \hspace{-1mm} {\alpha _m} \hspace{-1mm} \cdot \hspace{-1mm} L \hspace{-1mm} \cdot \hspace{-1mm} {SNR}_m} \right)}^2}}}
\end{align}

By letting~\eqref{df_SNR} be zero and utilizing \eqref{SNR}, the optimal conditions for each ${\alpha _m}$ can be obtained, as given by
\setcounter{equation}{11}
\begin{equation}\label{opt_con_SE}
    SNR = {\alpha _m^2 \cdot L \cdot SN{R_m} + 2 \cdot {\alpha _m} - 1},\ \forall m.
\end{equation}
By resolving \eqref{opt_con_SE}, two solutions will be derived, but only one is feasible due to ${\alpha _m}>0$, which is presented as
\begin{equation}\label{alpha_SE}
    {\alpha _m}{\rm{ = }}\frac{{{\theta _m} - 1}}{{L \cdot {SNR}_m}},
\end{equation}
where
\begin{equation}\nonumber
    \theta_m={\sqrt {1 + L \cdot {SNR}_m\left( {SNR + 1} \right)} }.
\end{equation}
Then substituting \eqref{alpha_SE} into \eqref{SNR}, the maximal $SNR$ must satisfy
\begin{equation}\label{SNR_Equ_SE}
    SNR = \frac{{\sum\limits_{m = 1}^M {\left( {1 - \frac{{{\theta _m} - 1}}{{L \cdot SN{R_m}}}} \right)SN{R_m}\frac{{{\theta _m} - 1}}{{{\theta _m}}}} }}{{\sum\limits_{m = 1}^M {\frac{{SN{R_m}}}{{{\theta _m}}}}  + M}}.
\end{equation}

Therefore, the maximal $SNR$, denoted by $SNR^*$, can be attained by solving \eqref{SNR_Equ_SE}. However, it is clear that the closed-form expression cannot be derived and only numerical result is available because of the complex structures of \eqref{SNR_Equ_SE}.

After computing the optimal $SNR^*$ and substituting it into \eqref{alpha_SE}, ${\alpha _m^*}$ can be obtained, as given by
\begin{equation}\label{alpha_opt_SE}
    \alpha _m^*=\frac{{\sqrt {1+L \cdot SN{R_m}\left( {SN{R^*} + 1} \right)}  - 1}}{{L \cdot SN{R_m}}}.
\end{equation}

However, only the numerical results are obtained in this way, so we further utilize some approximate relationships to acquire the analytic and closed-form solutions.

Without considering the arrangement of the signaling along the time and frequency, the total symbols during the channel coherence time interval can be roughly calculated by $L=B_c \cdot T_c$, where $B_c$ and $T_c$ are the channel coherence on frequency and time, respectively. And the typical value of $L$ and $B_c$ are $10^3$ and $370KHz$ in practice, based on which two approximate relationships can be obtained under high data rate demand, as written by
\begin{equation}\nonumber
    { {1 + L \cdot {SNR}_m\left( {SNR + 1} \right)} } \approx  {L \cdot {SNR}_m\left( {SNR + 1} \right)},
\end{equation}
\begin{equation}\nonumber
    {\frac{1}{{L\sqrt {L \cdot {SNR}_m} }}}  \approx 0.
\end{equation}

Utilizing the two approximate relationships above, \eqref{SNR_Equ_SE} can be transformed, as given by
\begin{equation}\label{SNR_appr_Equ_SE}
    M{\left( {\sqrt {SNR + 1} } \right)^2} + {b_{SE}} \cdot \sqrt {SNR + 1}  - {c_{SE}} = 0.
\end{equation}

Resolving \eqref{SNR_appr_Equ_SE} and considering $\sqrt {SNR + 1}>0$, ${SNR}^*$ can be derived, as given by
\begin{equation}\label{SNR_appr_opt_SE}
    {SNR}^* = {\left( {\frac{{ - {b_{SE}} + \sqrt {b_{SE}^2 + 4 \cdot M \cdot {c_{SE}}} }}{{2 \cdot M}}} \right)^2} - 1.
\end{equation}

Therefore, by substituting \eqref{SNR_appr_opt_SE} into \eqref{coh_AchiRate}, the maximal system capacity $C^*$ can be obtained and formulated as~\eqref{Maximum_Capacity}.
\end{IEEEproof}

\vspace{2mm}

And then the optimal pilot symbols ratio $\alpha _m^*$ can be calculated by substituting \eqref{SNR_appr_opt_SE} into \eqref{alpha_opt_SE}.
\section{EE Optimization}\label{sec:4}
In this section, we will discuss the optimal ratio of pilot symbols to maximize EE with required uplink data rate, denoted by $R_{ul}$. It is known that the total power consumption of the networks contains transmit power and circuit power. The circuit power can be further decomposed into static component (drive hardware) and dynamic component (process signal), which are denoted by $P_{base}$ and $P_c = \varepsilon \cdot R_{ul}$, where $\varepsilon$ is the power consumption for transmitting a data bit. The EE is defined as the radio between the average data rate and the average total power consumption at all nodes~\cite{Ref9}, as given by
\begin{equation}\label{EE_indicator}
{\eta _E} = \frac{R_{ul}}{P_{total}} = \frac{R_{ul}}{P + \varepsilon \cdot R_{ul} + P_{base}},
\end{equation}
which indicates that given the required uplink data rate, \emph{i.e.}, $R_{ul}$, maximizing $\eta_E$ is equivalent to minimizing the total transmit power $P$. Therefore, the EE optimization problem can be formulated, as given by
\begin{equation}\label{EE_Opt}
    \mathop {\min }\limits_{\{\alpha _m\}} \ P \tag{\textbf{P2}},
\end{equation}
\begin{equation}\nonumber
    \rm{s.t.}\ \ SNR \geq 2^{\frac{R_{ul}}{W}}-1,
\end{equation}
\begin{equation}\nonumber
    0 \le {\alpha _m} \le 1,\ m \in \{1, 2, \cdots, {M}\}.
\end{equation}
It can be easily proved that when the optimal solution is obtained, $SNR = 2^{\frac{R_{ul}}{W}}-1$ must be satisfied just by reducing transmit power $P$ to make it hold.

\vspace{2mm}

\begin{proposition} \label{EE_proposition}
\textit{With the required uplink data rate $R_{ul}$ in \mbox{non-coherent} \mbox{JR-CoMP} networks, the approximate minimal transmit power $P^*$ of the common UE can be presented by}
\begin{equation}\label{P_appr_opt_EE}
    {P^*} = {\left(\frac{{{b_{EE}} + \sqrt {b_{EE}^2 + 4 \cdot {c_{EE}} \cdot \sum\limits_{m = 1}^M {\frac{{{{\left| {{h_m}} \right|}^2}}}{{{\sigma ^2}}}} } }}{{2 \cdot \sum\limits_{m = 1}^M {\frac{{{{\left| {{h_m}} \right|}^2}}}{{{\sigma ^2}}}} }}\right)}^2,
\end{equation}
\textit{where}
\begin{equation}\nonumber
    {b_{EE}} = \sum\limits_{m = 1}^M {\frac{{2 \cdot \sqrt {\frac{{{{\left| {{h_m}} \right|}^2}}}{{{\sigma ^2}}} \cdot {2^{\frac{{{R_{ul}}}}{W}}}} }}{{\sqrt L }}},
\end{equation}
\begin{equation}\nonumber
    {c_{EE}} = M\left( {{2^{\frac{{{R_{ul}}}}{W}}} - 1} \right) - \frac{{2 \cdot M}}{L}.
\end{equation}
\end{proposition}


\begin{IEEEproof}
Lagrange multiplier method is adopted to prove this proposition, where the Lagrange multiplier is given by
\begin{equation}\label{Lagrange_multiplier}
    {\mathcal{L}_{EE}} = P - {\lambda _{EE}} \cdot \left( {SNR - {2^{\frac{{{R_{ul}}}}{W}}}+1} \right),
\end{equation}
where ${\lambda _{EE}}$ is the Lagrange coefficient. By letting $\frac{{{\partial ^2}{\mathcal{L}_{EE}}}}{{\partial \alpha _m^2}} = 0$ and utilizing \eqref{df_SNR}, the necessary condition for optimal $\alpha _m$ is
\begin{equation}\label{opt_con_EE}
    {\alpha _m^2 \cdot L \cdot SN{R_m} + 2 \cdot {\alpha _m} - 1} = {{2^{\frac{{{R_{ul}}}}{W}}}}.
\end{equation}

By solving \eqref{opt_con_EE} and considering ${\alpha _m}>0$, the optimal $\alpha _m$ can be expressed, as given by
\begin{equation}\label{alpha_opt_EE}
    \alpha _m=\frac{{\sqrt {1+L \cdot SN{R_m}{2^{\frac{{{R_{ul}}}}{W}}}}  - 1}}{{L \cdot SN{R_m}}}.
\end{equation}

By substituting \eqref{alpha_opt_EE} into \eqref{SNR} and considering $SNR={2^{\frac{{{R_{ul}}}}{W}}} - 1$, the following equation can be established as
\begin{equation}\label{P_Equ_EE}
    \frac{{\sum\limits_{m = 1}^M {\left( {1 - \frac{{\sqrt {{\vartheta _m}}  - 1}}{{L \cdot {SNR}_m}}} \right){SNR}_m\frac{{\sqrt {{\vartheta _m}}  - 1}}{{\sqrt {{\vartheta _m}} }}} }}{{\sum\limits_{m = 1}^M {\frac{{{SNR}_m}}{{\sqrt {{\vartheta _m}} }}}  + M}}={2^{\frac{{{R_{ul}}}}{W}}} - 1,
\end{equation}
where
\begin{equation}\nonumber
    {\vartheta _m} = 1+L \cdot {SNR}_m \cdot {2^{\frac{{{R_{ul}}}}{W}}}.
\end{equation}

Similarly, the optimal $P$, denoted by $P^*$, can be calculated by solving \eqref{P_Equ_EE}, based on which $\alpha _m^*$ can be obtained according to \eqref{alpha_opt_EE}. However, only the numerical results are acquired in this way due to the complex structures of \eqref{P_Equ_EE}.

Like SE maximization in the former section, some approximate relationships are introduced, as expressed by
\begin{equation}\nonumber
    {\vartheta _m} \gg 1,\ {\vartheta _m} \approx {\vartheta _m} - 1,
\end{equation}
\begin{equation}\nonumber
    {\frac{1}{{L\sqrt {L \cdot \frac{{{{\left| {{h_m}} \right|}^2}}}{{{\sigma ^2}}} \cdot {2^{\frac{{{R_{ul}}}}{W}}}} }}} \approx 0,
\end{equation}
based on which, \eqref{P_Equ_EE} can be transformed, as given by
\begin{equation}\label{P_appr_Equ_SE}
    P\sum\limits_{m = 1}^M {\frac{{{{\left| {{h_m}} \right|}^2}}}{{{\sigma ^2}}}}  - {b_{EE}}\sqrt P  - {c_{EE}} = 0.
\end{equation}

By solving \eqref{P_appr_Equ_SE} and considering $\sqrt{P}>0$, the optimal $P$ can be formulated, as given by~\eqref{P_appr_opt_EE}.
\end{IEEEproof}

\vspace{2mm}

And then the optimal pilot symbols ratio $\alpha _m^*$ can be obtained by substituting \eqref{P_appr_opt_EE} into \eqref{alpha_opt_EE}.
\section{Simulation results}\label{sec:5}
In this section, the typical scenario, involving three macro BSs, is considered in simulations to validate our optimal pilots assignment, where the parameters are specified in Table~I. Besides our schemes, some other schemes are contained in simulations for comparison, as described by
\begin{itemize}
  \item \textbf{Precise Optimization Scheme (POS)}: Obtain $\alpha_m^*$ and calculate ${SNR}^*$, $P^*$ by solving the equation precisely.
  \item \textbf{Approximate Optimization Scheme (AOS)}: Obtain $\alpha_m^*$, ${SNR}^*$, and $P^*$ by approximate expressions.
  \item \textbf{Genetic Algorithm Scheme (GAS)}: Obtain $\alpha_m^*$ by complicated genetic algorithm in Matlab.
  \item \textbf{Traditional Scheme (TA)}: Distribute the same $\alpha$ for each BS without optimization.
\end{itemize}

\vspace{-4mm}

\begin{table}[htbp]\label{Table_Simulation}
\centering
\caption{Simulation Parameters}
\vspace{-2mm}
\begin{tabular}{l|l}
\hline
\textbf{Parameters} & \textbf{Values} \\
\hline
System bandwidth ($W$) & 10\,MHz \\
Noise power spectral density ($N_0$) & --\,174\,dBm/Hz \\
Cooperative BSs number ($M$) & 3 \\
Distance between UE and BSs ($d_1,d_2,d_3$) & 200,250,300 \,m \\
Average path loss ($\text{PL}$) & $30+40\log_{10}d$ \,dB \\
Maximum transmit power ($P_{\max}$) & 46\,dBm \\
Static circuit power consumption ($P_{base}$) & 50\,mW \\
Dynamic circuit factor ($\varepsilon$) & 2\,mW/Mbps \\
The channel coherence interval ($L$) & $10^3$ \\
\hline
\end{tabular}
\vspace{-2mm}
\end{table}

\begin{figure}[t]
\centering
\hspace{-3mm}
\includegraphics[scale=0.25]{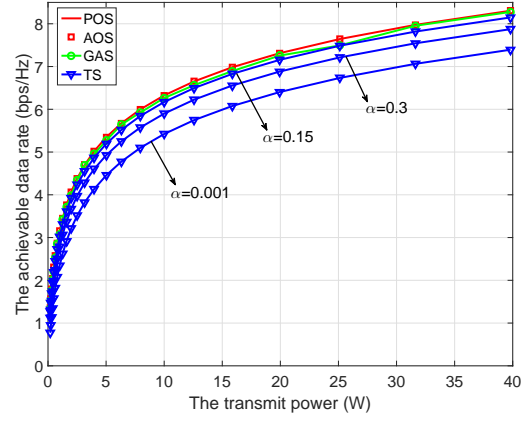}
\vspace{-3mm}
\caption{The achievable data rate region for different transmit powers.}
\label{SE}
\vspace{-5mm}
\end{figure}
\begin{figure}[t]
\centering
\hspace{-3mm}
\includegraphics[scale=0.25]{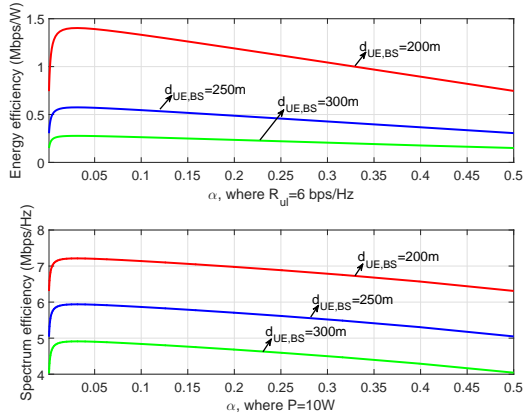}
\vspace{-3mm}
\caption{The effects of $\alpha$ on SE and EE respectively with different distance between UE and BSs.}
\label{alpha}
\vspace{-5mm}
\end{figure}

In Fig.\ref{SE}, it is observed clearly that POS can attain the highest achievable data rate for all transmit powers, indicating the validity and superiority of our scheme in terms of SE. The achievable data rates for all schemes will rise with the increment of transmit power $P$, but the growth rates all become lower because of the logarithmic relationship between the data rate and the SNR. It can be seen that AOS, much more simple than POS, can obtain the same SE performance, which will be more practical in reality. Therefore, the approximation is reasonable and quite precise. The randomness of GAS also can be found in the green curve, and GAS of high complexity can get the same SE compared with POS and AOS, implying POS and AOS are optimal. As for TS, different $\alpha$ will cause different performance, where overlow and overhigh $\alpha$ are both undesirable, which is carefully discussed in Fig.\ref{alpha}.

\begin{figure}[t]
\centering
\hspace{-3mm}
\includegraphics[scale=0.25]{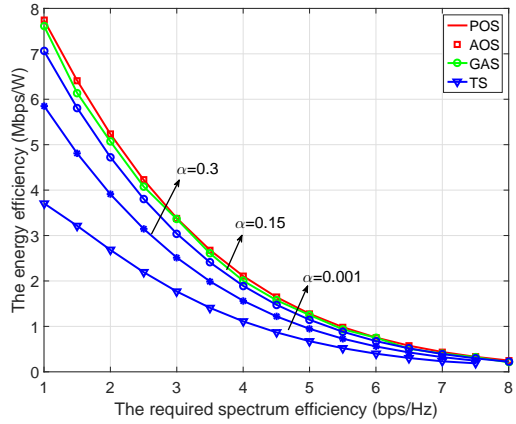}
\vspace{-3mm}
\caption{The optimal energy efficiency versus the required spectrum efficiency.}
\label{EE}
\vspace{-5mm}
\end{figure}
Fig.\ref{EE} compares the EE performance between different schemes with different required uplink data rate. It can be seen that POS can achieve optimal EE compared with TS and the performance of all schemes will be impaired with the increase of $R_{ul}$ due to the exponentially increasing nature of transmit power with respect to data rate growth. It is obvious that AOS can acquired the same EE performance with POS, revealing the rationality of our approximation. And the optimality of POS and AOS can be verified indirectly by GAS that almost have the identical EE performance. According to TS, suitable $\alpha$ needs to be chosen in terms of EE since different $\alpha$ will cause different performance like SE maximization.

In Fig.\ref{alpha}, it is seen that the performance first soars swiftly and then decreases gradually for both SE and EE. When $\alpha$ is very small, the channel estimation is extremely bad and the MMSE is high, which cause low useful signal power and much more severe interference. Therefore, the system performance is poor due to less correctly demodulated data even if there are so much resource for desired data. Under this region, called the channel estimation limited region, a little growth of $\alpha$ will improve the channel estimation considerably and obtain a performance boost due to the property of MMSE (inverse function w.r.t $\alpha$). On the contrary, when $\alpha$ is large enough, the channel estimation improves very slow due to the inverse function in MMSE. If we continue to increase $\alpha$, less symbols will be left and the performance will deteriorate. This region is called useful information limited region.
\section{Conclusion}\label{sec:6}
In this paper, with the consideration of channel estimation error and non-ideal backhaul links, the optimal pilot symbols ratio is derived to maximize SE and EE in uplink JR-CoMP networks. The maximal system capacity in SE optimization and the minimal transmit power in EE optimization are also obtained by solving complicated equations. In order to reduce the complexity, some reasonable approximations are introduced to obtain the closed-form expressions of these results. Simulation reveals the superiority of our scheme, the accuracy of our approximations, and the effect of pilot symbols ratio.
\section*{Acknowledgement}
The work was supported by National Nature Science Foundation of China Project (Grant No.61471058), Hong Kong, Macao and Taiwan Science and Technology Cooperation Projects (2014DFT10320, 2016YFE0122900), the 111 Project of China (B16006)​ and Beijing Training Project for The Leading Talents in S\&T (No. Z141101001514026).


\begin{thebibliography}{13}
\bibitem{Ref9}
Q. Cui, T. Yuan, and W. Ni, ``Energy-efficient two-way relaying under non-ideal power amplifiers,'' \textit{IEEE Trans. Veh. Technol.}, vol. 66, no. 2, pp. 1257-1270, 2017.
\bibitem{Ref4}
M. Cai, and J. N. Laneman, ``Wideband distributed spectrum sharing with multichannel immediate multiple access,'' \textit{Analog Integr. Circ. and Signal Process}, Springer, https://arxiv.org/pdf/1702.02695.pdf.
\bibitem{Ref12}
O. Onireti, F. Heliot, and M. A. Imran, ``On the energy efficiency-spectral efficiency trade-off in the uplink of CoMP system,'' \textit{IEEE Trans. Wireless Commun.}, vol. 12, no. 2, pp. 556-561, 2012.
\bibitem{Ref10}
Q. Cui, H. Song, and H. Wang \textit{et al.}, ``Capacity analysis of joint transmission CoMP with adaptive modulation,'' \textit{IEEE Trans. Veh. Technol.}, no. 99, pp. 1-1, 2016.
\bibitem{Ref2}
S. Sun, Q Gao, Y Peng \textit{et al.}, ``Interference management through CoMP in 3GPP LTE-advanced networks,'' \textit{IEEE Wireless Commun.}, vol. 20, no. 1, pp. 59-66, 2013.
\bibitem{Ref8}
Q. Cui, H. Wang, and P. Hu \textit{et al.}, ``Evolution of limited-feedback CoMP systems from 4G to 5G: CoMP features and limited-feedback approaches,'' \textit{IEEE Veh. Technol. Mag.}, vol. 9, no. 3, pp. 94-103, 2014.
\bibitem{Ref13}
P. Marsch and G. Fettweis, ``Uplink CoMP under a constrained backhaul and imperfect channel knowledge,'' \textit{IEEE Trans. Wireless Commun.}, vol. 10, no. 6, pp. 1730-1742, 2011.
\bibitem{Ref14}
Y. Cai, F. R.Yu and G. Senarath, ``Optimal clustering and rate allocation for uplink coordinated multi-point (CoMP) systems with delayed channel state information (CSI),'' \textit{in Proc. ICC}, pp. 6025-6029, 2013.
\bibitem{ChannelError}
B. Hassibi and B. M. Hochwald, ``How much training is needed in multiple-antenna wireless links?,'' \textit{IEEE Trans. Inf. Theory}, vol. 49, no. 4, pp. 951-963, 2003.
\bibitem{Ref7}
M. Karami, A. Olfat, and N. C. Beaulieu, ``Pilot symbol parameter optimization based on imperfect channel state prediction for OFDM systems,'' \textit{IEEE Trans. Commun.}, vol. 61, no. 61, pp. 2557-2567, 2013.
\bibitem{Ref3}
R. Nissel, and M. Rupp, ``Bit error probability for pilot-symbol aided channel estimation in FBMC-OQAM,'' \textit{in Proc. ICC}, pp. 1-6, 2016.
\bibitem{Ref15}
K. Takeda, Y. Yuda, and M. Hoshino \textit{et al.}, ``Impact of channel estimation error on LTE-advanced uplink using CoMP joint reception in a heterogeneous network,'' \textit{in Proc. VTC-Spring}, pp. 1-5, 2013.

\end{thebibliography}
\end{document}